\documentclass[aps,prl,reprint,longbibliography,superscriptaddress,noeprint]{revtex4-2}
\usepackage{lipsum}
\usepackage{siunitx}
\usepackage{amsmath,bm}
\usepackage{amsbsy}
\usepackage{amssymb}
\usepackage{textcomp}
\usepackage{color}
\usepackage{braket}
\usepackage{graphicx}
\usepackage{textcomp}
\usepackage{notes2bib}
\usepackage{textcomp}
\usepackage{mwe}    
\usepackage[normalem]{ulem}
\usepackage{siunitx}
\usepackage{filecontents}
\usepackage{soul}

\usepackage{svg}

\definecolor{bl}{rgb}{0, .1, .6}
\definecolor{gr}{rgb}{.2, .6, .2}
\usepackage[colorlinks=true, citecolor = bl, linkcolor = bl, urlcolor=bl, pdfborder={0 0 0}]{hyperref}

\begin{document}
\title{Fourier imaging of collective spontaneous emission modes in superradiant cold atomic clouds}

\author{Adrien Gavalda}\altaffiliation{These authors contributed equally to this work.}
\affiliation{Universit\'e Paris-Saclay, Institut d'Optique Graduate School, CNRS, 
Laboratoire Charles Fabry, 91127, Palaiseau, France}
\affiliation{PASQAL SAS, 24 rue Emile Baudot, 91120 Palaiseau, France}
\author{Guillaume Tremblier}\altaffiliation{These authors contributed equally to this work.}
\author{Martin Poitrinal}\altaffiliation{These authors contributed equally to this work.}
\author{Sara Pancaldi}
\author{Antoine Browaeys}\email{antoine.browaeys@institutoptique.fr}
\author{Igor Ferrier-Barbut}\email{igor.ferrier-barbut@institutoptique.fr}

\affiliation{Universit\'e Paris-Saclay, Institut d'Optique Graduate School, CNRS, 
Laboratoire Charles Fabry, 91127, Palaiseau, France}

\begin{abstract}
We measure the spatial pattern associated with the superradiant emission from a cloud of cold $^{87}$Rb atoms using Fourier imaging. We observe  a highly directional, ring-shaped emission structure, which corresponds to the electromagnetic mode emitted by the most superradiant collective jump operator. 
	Using spatial filtering, we isolate this channel and find the typical superradiant burst with superlinear scaling of the intensity with atom number. 
We compare our results to two models that describe the competition between the various decay channels, finding good agreement. 
Our work shows that the collective jump operators introduced by Carmichael {\it et al.} \cite{Carmmichael2000} can be measured and manipulated.
\end{abstract} 

\maketitle

Dicke predicted  that an ensemble of identical two-level atoms, all located within a sub-wavelength sized volume and initially in their excited state, decays by emitting a burst of so-called superradiant radiation \cite{Dicke1954}. 
In his scenario, the atoms are indistinguishable and collectively coupled to a single channel of spontaneous emission.
The burst builds up thanks to the enhancement of the emission rate in this decay channel. 
The first observations of Dicke superradiance  
were however performed on pencil-shaped samples larger than the transition wavelength, the emission occurring along the main axis \cite{Skribanowitz1973,Gross1976,Heinzen1985,Inouye1999,Lopes2014}. 
These observations were explained by a mean-field treatment using the Maxwell-Bloch equations, relying on an atomic spin-wave written along the main axis after the emission of the first photons:  
the dynamics then follows closely that of Dicke, replacing the single atom decay rate $\Gamma_0$ by a collective one $\mu\Gamma_0$, where $\mu$ is the cloud form factor \cite{Gross1982,Allen1987}. 
As a consequence, superradiance occurs above a critical  atom number $N_c\sim1/\mu$.
However the spin-wave ansatz still restricts the evolution of the system to a single decay channel, 
while the breaking of indistinguishibility of the atoms due to the finite sample size should involve many channels for collective spontaneous emission. 

To explore this question, new generation experiments in the optical domain, using cold atoms \cite{Norcia2016,Ferioli2021a,Liedl2024,Glicenstein2024,Douglas2026}, were paralleled by intense theoretical investigations of the role of the geometry and size of the ensembles on collective spontaneous emission \cite{Carmmichael2000,Clemens2003,Sutherland2017,Masson2020,Robicheaux2021,Masson2022,Holzinger2025,Rusconi2026}.
In this context, Carmichael {\it et al.} \cite{Carmmichael2000,Clemens2003} introduced the concept of collective jump operators  to describe the superradiant emission by finite size ensembles. 
These operators, and their associated electromagnetic (EM) modes, could be useful concepts to devise efficient light-matter interfaces based on free-space atomic ensembles  \cite{Facchinetti2016,Rui2020,Solomons2025,N-atom-paper}. 
Here we demonstrate experimentally that, although these collective atomic modes have so far remained a theoretical tool,
they can in fact be measured and isolated. 
To do so, we record the spatial pattern of the  light emitted by an elongated atomic sample using Fourier imaging, revealing in particular the most superradiant EM mode.
Furthermore, using spatial filtering, we separate the contributions of various EM modes in the emission and explain our observations by a condition for superradiance to occur in a given EM mode. We also compare our findings to two models describing the competition between the various atomic modes, finding good agreement. 

\begin{figure*}[t!]
    \centering
     \includegraphics[width=\textwidth]{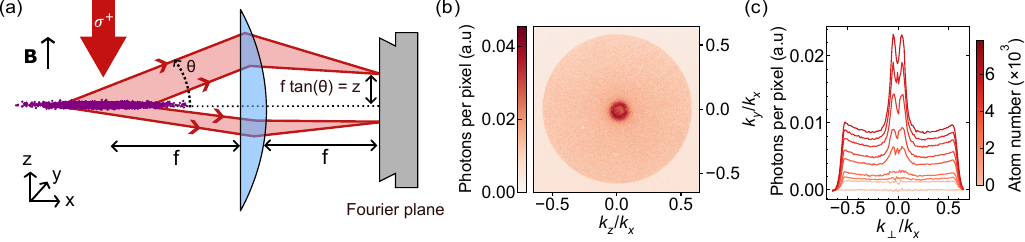}
    \caption{(a) Simplified schematics of the experiment. The light emitted by an initially inverted elongated cloud of two-level Rb atoms is collected on an EMCCD camera placed in the Fourier plane of the imaging system: the position in this plane is related to the wavevector components by $(y,z) =(k_y,k_z)f/k_x$ with $\tan\theta=k_{y,z}/k_x$. Effective focal length $f$ is equal to 21 mm.
    (b) Fourier plane image of the emission pattern of a cloud containing $N= 5000$ atoms, integrated over a superradiant burst. 
    (c) One-dimensional azimuthal averages of the Fourier plane image for $N= (0, 750, 1500, 2200, 3900, 5000, 6000, 7100, 7400)$.}  
    \label{fig:setup}
\end{figure*}

Our experimental setup \cite{Pancaldi2025} relies on a glass cell, surrounded by three microscope objectives with numerical aperture ${\rm NA}=0.5$ used to focus an optical dipole trap (ODT) and collect the light emitted by the atoms.
It produces clouds of \textsuperscript{87}Rb atoms loaded in the ODT from a magneto-optical trap using grey molasses on the $D_1$ transition \cite{Glicenstein2021}. The ODT ($\lambda=$ \SI{935}{nm}, power up to $\SI{1}{W}$) 
has a waist radius controlled by tunable lenses between $w_0=\SI{2.5}{\micro m}$ and $\SI{8}{\micro m}$.
The resulting elongated clouds contain up to $N=8000$ atoms, at a temperature of $T=550-\SI{1550}{\micro K}$, with {\it calculated } Gaussian rms axial (radial) sizes ranging from $\sigma_x = \SI{23}{\micro m}$ to $\SI{80}{\micro m}$ ($\sigma_r=\SI{1.3}{\micro m}$ to $\SI{2.2}{\micro m}$). The systematic relative uncertainties in the determination of the sizes are 25\% (axial) and 20\% (radial).
We apply a magnetic field of {\SI{64}{G} perpendicular to the cloud axis [see Fig.\,\ref{fig:setup}(a)] to isolate a closed $\sigma^+$-transition between the two states of the D2 line $\ket g=\ket{5S_{1/2},F=2, m_F=2}$ and $\ket{e}=\ket{5P_{3/2},F'=3,m'_F=3}$ ($\lambda_0=2\pi/k_0 \approx \SI{780.2}{\nano\meter}$, $\Gamma_0 \approx 2\pi \times \SI{6.1}{\mega\hertz}$ and $I_{\text{sat}}\simeq \SI{1.67}{\milli\watt/\centi\meter^2} $).

During the experimental sequences, the trap is turned off and the atoms are excited by a laser beam propagating perpendicularly to the cloud's main axis, linearly polarized along $x$ (superposition of $\sigma^+$ and $\sigma^-$ polarizations). 
Its frequency is resonant with the $\sigma^+$-transition $\ket e\to\ket g$, the $\sigma^-$ component being detuned by \SI{119}{MHz} from the $m_F=2\to m'_F=1$ transition, thus playing no role.
We reach  Rabi frequencies of up to $\Omega=2\pi\times \SI{55}{MHz} \simeq 9\Gamma_0$ and apply a $\pi$-pulse of duration $0.34/\Gamma_0= 9$\,ns to invert the atomic population, yielding an initial excited state population of $88\%$ limited by the residual decay during excitation \cite{Steck}, and no initial coherence along the cloud axis \cite{Ferioli2021a}. After the atoms have decayed, we recapture the cloud and repeat the experiment 200 times on the same  cloud ensuring the atom number remains constant within 20\%.

Using one of the two axial objectives, we image the emitted light in Fourier space by conjugating the back focal plane of the objective with an EMCCD camera [Fig.\,\ref{fig:setup}(a), relay telescope not shown]. 
We collect the wavevectors $\mathbf k$ with $\theta = ({\bf e}_x,{\bf k})\leqslant\sin^{-1}({\rm NA})=30^\circ$. 
The dipolar radiation pattern $I_0(\mathbf k)\propto 1+(\mathbf{{\hat{k}}} \cdot\mathbf e_z)^2$ ($\mathbf{k}=k_0\mathbf{\hat{k}}$) is nearly homogeneous over the collection angle.
Figure \ref{fig:setup}(b) shows the recorded Fourier plane image resulting from the integration of a superradiant burst of $N=5000$ atoms, averaged over 40000 identically prepared clouds. 
We observe two contributions: first, a homogeneous background covering the full aperture; 
second, a highly directional, ring-shaped emission along the $x$ axis.
To verify that this last contribution originates from superradiance, we show in Fig.\,\ref{fig:setup}(c) 
the azimutally-averaged images for various
atom numbers: the directional emission occurs above a critical atom number $N_c$ and increases nonlinearly with atom number, as expected for superradiance.
Furthermore, the width and shape of the emission pattern does not evolve with atom number above $N_c$, suggesting that it depends solely on the cloud geometry.
The pattern is rotationally symmetric around the cloud axis but, strikingly, is not peaked at $\theta = 0$. 
This is in stark contrast with the prediction from the standard theory of superradiance \cite{Gross1982,Allen1987}, which assumes that a spin-wave with wavevector $|k|=k_0$ forms along the cloud axis: this would result in an emission pattern peaking at $\theta=0$. 

To explain our observations, we use the framework introduced by Carmichael {\it et al.}~to describe the collective spontaneous emission of $N$ two-level atoms~\cite{Carmmichael2000,Clemens2003}. 
It relies on the master equation $\dot \rho= 
{\cal L}[\rho]$  with the non-diagonal Lindbladian given by ${\cal L}[\rho]=\sum_{ij}\Gamma_{ij}(\sigma_j^-\rho\sigma_i^+-\{\sigma_i^+\sigma_j^-,\rho\}/2)$ \cite{Lehmberg1970}. 
Here, $\Gamma_{ij}=\frac{6\pi\Gamma_0}{k_0} \text{Im}[G\left(\mathbf{r}_i-\mathbf{r}_j\right)]$, where $G({\bf r})$ is the EM vacuum Green's function \cite{SM}.
The diagonalization of the $\Gamma_{ij}$ matrix leads to $N$ collective jump operators (which we call atomic modes)  $S_l^-=\sum_i\alpha_{il}\sigma^-_i$ ($\alpha_{il}\in \mathbb{R}$) associated to the collective rates $\Gamma_l$.
In this basis, the Lindbladian reads ${\cal L}[\rho]=\sum_{l}\Gamma_{l}(S_l^-\rho S_l^+-\{S_l^+S_l^-,\rho\}/2)$.
The radiation pattern of each atomic mode in the far field is $E_l(\mathbf k)\propto \sum_i\alpha_{il}e^{i\mathbf{k}\cdot\mathbf{r}_i}$, yielding an intensity pattern 
\begin{equation}
I_l(\mathbf{k}) = I_0(\mathbf{k})\sum_{ij}\alpha_{il}\alpha_{jl}e^{i\mathbf{k}\cdot(\mathbf{r}_i-\mathbf{r}_j)},\label{eq:modepattern} 
\end{equation}
with $I_0(\mathbf{k})$ the single-atom dipolar emission pattern of the considered transition. The $S_l^-$ operators thus describe the emission of one photon in the EM mode $I_l$~\cite{N-atom-paper}.

To compare these patterns with our observations, we compute the decay matrix $\Gamma_{ij}$ and diagonalize it for cloud sizes $\sigma_x = \SI{78}{\micro m}$ and $\sigma_r =\SI{1.7}{\micro m}$, and various atom numbers. 
Figure\,\ref{fig:modes}(a) shows the distribution of eigenvalues $\Gamma_l/\Gamma_0$, featuring atomic modes with $\Gamma_l>\Gamma_0$ and $\Gamma_l<\Gamma_0$. As expected, the distribution is broader for the largest atom number. 
In addition, a pair of nearly degenerate superradiant modes emerges associated to $\Gamma_1$ and $\Gamma_2$, separated by a gap from the bulk of eigenvalues. They correspond to two standing waves along the cloud axis dephased by $\pi/2$ \cite{SM}.
We plot in Fig.\,\ref{fig:modes}(b) the angular emission pattern of that pair 
calculated using Eq.\,\eqref{eq:modepattern}, and compare its azimuthal average to the measured one in Fig.\,\ref{fig:modes}(c). 
Only the most superradiant pair covers a small angular aperture and agrees well with the experimentally observed central feature \cite{SM}.
This suggests that it represents the direct observation of the action of the collective jump operator associated to the most superradiant pair. 
The background observed on the data results from the contributions of the emission patterns from the other atomic modes which are less 
and less directional as their index increases. Examples are shown in \cite{SM}.

\begin{figure}
    \centering
    \includegraphics[width=1\columnwidth]{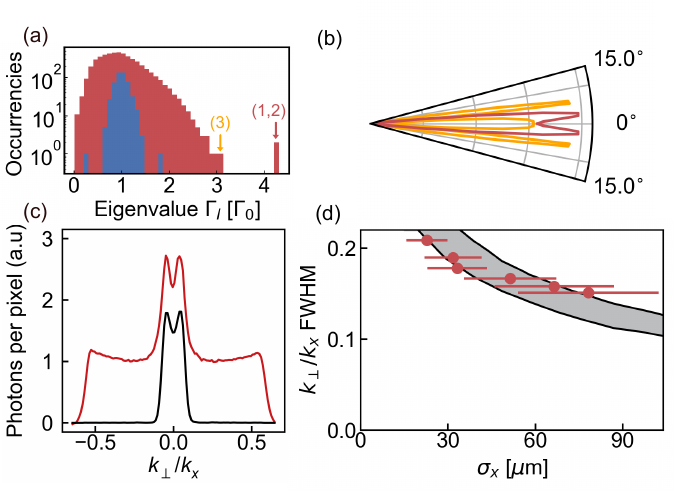}
    \caption{Collective modes of a cloud with $\sigma_x = 78\,\mu{\rm m}$ and $\sigma_r =1.7\,\mu{\rm m}$. 
    (a) Distribution of $\Gamma_{l}$'s  for $N=5000$ atoms (red) and $N=500$ atoms (blue).
    (b) Normalized angular emission pattern for the most superradiant {atomic modes 1 and 2} (red), and the third mode (orange). 
    (c) Azimuthal average of the emission pattern comparing experiment in red ($\sigma_x =78(20)\,\mu{\rm m}$ and $\sigma_r =2.2(5)\,\mu{\rm m}$) and the theoretical pattern of the most superradiant atomic mode (black) with an arbitrary vertical scaling factor. 
    (d) FWHM of the superradiant emission pattern (experiment in red) and the corresponding theoretical values (in black) as a function of rms length $\sigma_x$ of the cloud, including its coupling to $\sigma_r$. The error region (gray area) in the theoretical prediction accounts for the experimental uncertainty on the cloud's radius $\sigma_r$.
    }
    \label{fig:modes}
\end{figure}

To verify that the spatial pattern of the most superradiant pair depends on the cloud geometry, we vary experimentally its dimensions by changing the dipole trap waist. 
We plot in Fig.\,\ref{fig:modes}(d) the FWHM of the measured intensity pattern and compare it to the theoretical one for various cloud lengths, finding good agreement. 
Finally, we have numerically found  that the specific shape of the emission pattern of the most superradiant atomic mode, not peaked at $\theta=0$, results from the specific dimensions of our clouds and is a compromise between a phase matching condition along the cloud axis and the finite transverse size of the clouds \cite{SM}.

\begin{figure}[b]
    \centering
    \includegraphics[width=1\columnwidth]{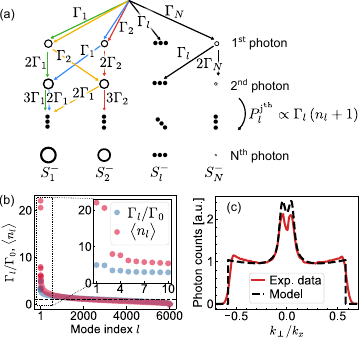}
    \caption{(a) Schematic representation of superradiant cascade in the bosonic model (see text). 
    (b) Calculated number of photons emitted per mode using the bosonic model, compared to the eigenvalues $\Gamma_l/\Gamma_0$ for a cloud with $\sigma_x=78\,\mu{\rm m}$, $\sigma_r=1.7\,\mu{\rm m}$.
    (c) Calculated emission pattern integrated over the duration of the decay (dotted line), compared to the experimental data.}
    \label{fig:cascade}
\end{figure}

The agreement between the experimental emission pattern and the one of the most superradiant pair raises the question of why it dominates despite the fact that a large fraction of the  $N$ modes also
have $\Gamma_l>\Gamma_0$.
To explain this, we now study the atomic mode competition during the decay.
After full atomic inversion, the probability that the first photon is emitted by atomic mode $l$ is $\Gamma_l/(N\Gamma_0)$, with pattern $I_l(\mathbf{k})$.
Summing the weighted patterns over all modes gives the dipolar pattern \cite{Clemens2003} in which the first photon is thus emitted.
At later times, the instantaneous emission rate by atomic mode $l$ is given by $\Gamma_l\langle S_l^+S_l^-\rangle(t)$. 
Emitting a photon by atomic mode $l$ reinforces these atomic correlations, increasing the probability of emission of the following photons by the same mode \cite{Masson2020}.
As a consequence, jump operators with the largest decay rate $\Gamma_l$ take over as the decay proceeds. 
A full calculation requires solving the master equation, out of reach for our atom numbers. 
To obtain an intuitive picture, we use the bosonic approximation of the master equation introduced in~\cite{Nowak2008}: 
the model replaces the collective jump operators $S_l^-$ acting on the initial inverted state $\ket{ee\cdots e}$ by a bosonic creation operator $a_l^\dagger$ acting on vacuum $\ket0$.
The fact that the ensemble hosts at most $N$ excitations is enforced by adding an annihilation operator $b$, where $b^\dagger b$ counts the number of excited atoms, leading to the mapping $S_l^-\to a_l^\dagger b$. 
This replacement gives an emission rate for atomic mode $l$ after $k$ jumps, $R_l^{(k)}(n_l)=\Gamma_l(1-\frac{k}{N})(n_l+1)$, with $n_l$ the number of photons already emitted by that mode: the approach thus 
describes the amplification of the emission rate as a bosonic enhancement akin to stimulated emission.
The scenario for superradiant emission is now the following: 
the photon emissions follows a trajectory on a graph, as represented in Fig.\,\ref{fig:cascade}(a), 
where the most superradiant pair is amplified more than the others and rapidly dominate. 
Figure\,\ref{fig:cascade}(b) compares the average number $\langle n_l\rangle$ of photons emitted by atomic mode $l$ during the full decay of $N$ excited atoms, to $N$ times the probability of emitting \emph{the first photon} by atomic mode $l$, $\Gamma_l/\Gamma_0$.
It confirms that the distribution of photon emissions by the different modes skews towards the most superradiant ones. 
However, we find it crucial for the most superradiant pair to stand out that the distribution of eigenvalues features a gap: in this case, the ratio of the number of photons emitted in the most superradiant pair with respect to the one emitted by the next atomic modes gets amplified, as illustrated in Fig.\,\ref{fig:cascade}(b).  
Finally, we calculate the total emission pattern by summing a repetition of 2000 simulated trajectories, and compare it to the measured one \footnote{We fix the scale of the theory so that the value of the background radiation matches the experiment.}: the result is shown in Fig\,\ref{fig:cascade}(c). 
The agreement between the experiment and the theory (shape and amplitude of the central peak) is remarkable, despite the simplifying assumptions
of the bosonic mode approach.

To further demonstrate that the collective jump operators provide the proper representation of collective spontaneous emission in experiments, we turn to the time-dependence of the emitted light during the superradiant decay.
For this, we instead use avalanche photodiodes in single photon counting mode, coupled to single-mode fibers. 
We select the collected wavevectors using masks in a Fourier plane (back focal plane of the objective). 
We first collect the light  along the cloud axis with one objective, blocking all wavevectors that are out of the peaked, superradiant emission. 
The results are shown in Fig.\,\ref{fig:dynamics}(a) and exhibit a burst characteristic of Dicke superradiance. 
To verify its collective character, we plot in Fig.\,\ref{fig:dynamics}(b) the maximum emission rate per atom normalized to its initial value as a function of the atom number. 
We observe a superradiant burst above a critical atom number $N_c$, with the maximum of the emission rate occurring after the extinction of the driving pulse.
To explain the existence of a critical atom number, we follow Ref.\,\cite{Robicheaux2021} and calculate the derivative of the initial emission rate in EM mode $l$, finding: $\dot I_l(0)=(\Gamma_l-2\Gamma_0)\Gamma_l$ \cite{SM}. 
Consequently, an atomic mode with $\Gamma_l> 2\Gamma_0$ leads to the emission of a burst rather than to an exponential decay. 
We hence recover the mean-field theory prediction of a critical atom number, but now \emph{per mode}, $N_{{\rm c},l}$ defined by $\Gamma_l(N_{{\rm c},l})=2\Gamma_0$.
For our cloud sizes, we find numerically that $\Gamma_1$ scales linearly with $N$, $\Gamma_1\simeq (\eta N+1)\Gamma_0$ with $\eta\approx 6\times10^{-4}$.
This leads to $N_{\rm{c},1}=1/\eta\approx1700$. 
We also find that the peak emission rate scales as $(N-N_{\rm{c},1})N\approx N^2$ for $N\gg N_{\rm{c},1}$, compatible with a Dicke scenario.

Next,  we repeat the same experiment, blocking the wavevectors emitted within the most superradiant EM mode, thus collecting the emission from atomic modes $l\geqslant 3$ whose emission patterns have a large angular width [Fig.\,\ref{fig:modes}(b)]  \footnote{In our previous study \cite{Ferioli2021a}, we collected the light in an aperture contained both the center and surrounding region, thus averaging over several superradiant modes.}. 
We still observe a non-exponential decay.
This is expected, as many modes $l\geqslant3$ also satisfy the condition $\Gamma_l>2\Gamma_0$ for superradiant emission.
Finally, we collect the light perpendicularly to the cloud axis, a case for which we expect to measure the emission from many atomic modes with $\Gamma_l\lesssim \Gamma_0$.
The intensity decays nearly exponentially, with a characteristic time close to single atom decay although slightly faster for high atom number. For  all atom numbers, we observe no initial increase of the emission rate (no burst)  [Fig.\,\ref{fig:dynamics}(b)].

\begin{figure}[t]
    \centering
    \includegraphics[width=1\columnwidth]{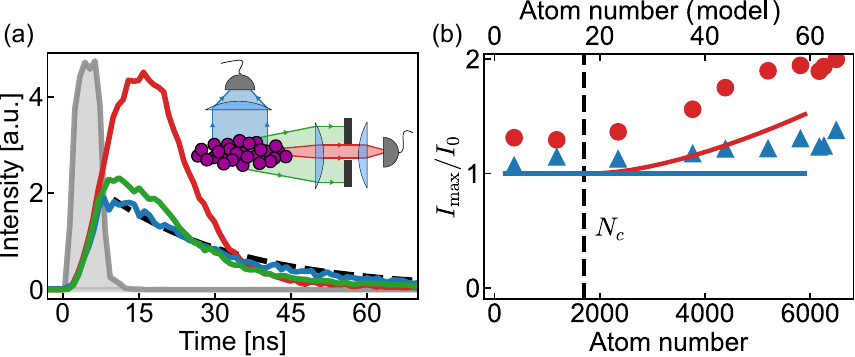}
    \caption{(a) Superradiant temporal emission measured after a $\pi$-pulse ($N = 6300$), with various collection schemes.   
    Red: selection of the most superradiant mode. 
    Blue: direction orthogonal to the cloud's axis. 
    Green: Axial emission, blocking the superradiant light.
    Dashed line: single atom decay
    (b) Maximum intensity $I_{\rm max}$, relative to the intensity $I_0$ at $t=\SI{6}{ns}$, a time chosen because at high atom number superradiance begins before the pulse end ($\SI{9}{ns}$).
    Solid lines: minimalistic model (see text). Dashed vertical line:  theoretical critical atom number for both cases as defined in the text. Colors are identical to panel (a).
    }
    \label{fig:dynamics}
\end{figure}
While the bosonic model explains the amplifications of the most superradiant pair, it does not reproduce the exact time dependencies observed, neither predicts a critical atom number (see \cite{SM}).
We thus introduce a second model that keeps the collective jump description of the ensemble. It isolates one collective atomic mode, and adds other atomic  modes as individual decays \cite{Leppenen2025}.
Its Lindblad operator is $\mathcal L[\rho]=\sum_{i}\Gamma_{0}(\sigma_i^-\rho\sigma_i^+-\{\sigma_i^+\sigma_i^-,\rho\}/2)+N\Gamma_{c}(S^-\rho S^+-\{S^+S^-,\rho\}/2)$, with $S^-=\frac{1}{\sqrt N}\sum_i\sigma^-_i$ and $\Gamma_c/\Gamma_0=\eta_*<1$. 
This model is permutationally symmetric and can thus be efficiently simulated for tens of atoms using the Qutip and PIQS frameworks \cite{Lambert2026,Shammah2018}. 
It corresponds to $\Gamma_{ij}=\Gamma_c+\delta_{ij}\Gamma_0$, with eigenvalues $\Gamma_1=N\Gamma_c+\Gamma_0$, $\Gamma_{2\cdots N}=\Gamma_0$.
We plot in Fig.\,\ref{fig:dynamics}(b) the peak emission rates for the individual and collective channels: $\langle\sigma_i^+\sigma_i^-\rangle_{\rm max}$, $\langle S^+S^-\rangle_{\rm max}$, as well as the peak total radiated power, as a function of atom number, taking $\eta_* = 0.06$ \footnote{$\eta_*$ is chosen so that $N_c$ does not exceed our computational power}. 
This minimalistic model reproduces the  experimental observations: the collective emission rate is $\propto N$ below a threshold $N_{\rm{c},1}= 1/\eta_*\approx17$ before Dicke superradiance sets in and then scales as $N(N-N_{\rm{c},1})$. 

Our work suggests several directions of research. 
First, the  gap between the most superradiant mode and the bulk of eigenvalues seems necessary for it to dominate the dynamics. 
We observed that the existence of this gap depends of the cloud geometry, but could not find an explanation of this fact in the context of random matrix theory \cite{Skipetrov2011,Viggiano2023}. 
Second,  to describe our observations and get physical intuition we have employed simplified models, which ignore in particular the hamiltonian part of the dynamics coming from the resonant dipole interactions \cite{SM}.
It would be interesting to explore its role using numerical methods such as the cumulant expansion or the truncated Wigner approximation \cite{Ferioli2021a,Mink2023}. 
Last, our demonstration that collective spontaneous emission modes can be manipulated will be useful for future experiments aiming at light-matter interfaces made of arrays of quantum emitters \cite{Facchinetti2016,Bettles2016,Shahmoon2017,N-atom-paper,Asenjo2017,Manzoni2018,Bekenstein2020,Rui2020,Srakaew2023,Solomons2025}. 
For example, one could  use a spatial light modulator to excite precisely through one EM mode or selectively interfere reference light with it.

\begin{acknowledgments}
We thank Ana Asenjo-Garcia and Ephraim Shahmoon for discussions, as well as Jean-René Rullier for assistance in the design of the experiment. We acknowledge useful discussions with the members of the PANDA collaboration.   
This project has received funding by the Agence Nationale de la Recherche (ANR-22-PETQ-0004 France 2030, project QuBitAF), by the European Union (ERC AdG ATARAXIA 101018511), from the EU’s Horizon Europe program under grant agreement No. 101115420  (PANDA project) and the Horizon Europe program HORIZON-CL4- 2022-QUANTUM-02-SGA (project 101113690 PASQuanS2.1).  
S.~P.~ and G.~T.~acknowledge funding by by the Paris Saclay Quantum Center.\\
All of the data from this manuscript is available on Zenodo \cite{zenodo}.
\end{acknowledgments}

\bibliography{biblio}

\clearpage
\newpage
\begin{center}
	{\Large Supplemental Material}
\end{center} 
\setcounter{figure}{0}
\renewcommand\thefigure{S\arabic{figure}} 
\setcounter{equation}{0}
\renewcommand\theequation{S\arabic{equation}} 

\section{Expression of the vacuum Green's function}

For our experimental case of atoms with two levels coupled by a $\sigma_+$-transition, the expression of the Green tensor $G(\mathbf{r})$ of the electromagnetic field in vacuum, projected on the polarization $\mathbf{p}\propto-( {\bf e}_x+i{\bf e}_y)$ is \cite{Asenjo2017}:
\begin{eqnarray}
    G(\mathbf{r}) = \mathbf{p}^*.\bar{\bar{G}}(\mathbf{r}).\mathbf{p} = \frac{k_0}{4\pi} e^{ik_0 \mathbf{r}} \times \biggl\{ \left[ 1-|\hat{\mathbf{p}}.\hat{\mathbf{r}}|^2 \right] \frac{1}{kr} \nonumber \\ + \left[3|\hat{\mathbf{p}}.\hat{\mathbf{r}}|^2 - 1\right] \left(  \frac{1}{(kr)^3} - \frac{i}{(kr)^2} \right)
    \biggr\}  
\end{eqnarray}
where $\hat{\mathbf{r}}$ is a normalized vector pointing in direction $\mathbf{r}.$

\section{Calibration of the atom number}

We estimate the atom number by calibrating the collection efficiency of our imaging system using a single atom loaded in the tightest trap we can produce with our tunable telescope ($w=1.75~\si{\micro\meter}$). 
We  image it for $10\,\mu$s with a single resonant beam well above saturation so that the scattering rate is $\Gamma/2\approx 2\times 10^{7}$\,s$^{-1}$. 
This corresponds to 190 scattered photons, to be compared to an average of $9$ photons measured on the EMCCD camera. 
The collection efficiency of our imaging system is thus $4.7\%$. 
To measure the atom number in a $N$-atom clouds, we turn off the trap for 5 $\mu$s to obtain a dilute sample in which collective effects are negligible. We then apply the same single resonant beam at $I/I_{\rm sat}\gg1$ for $10\,\mu{\rm s}$ and count the number of detected photons which we divide by the value for a single atom to obtain the number of atoms. 

\section{EM Modes associated to the collective jump operators}\label{SM:modes}

\begin{figure}
    \centering
    \includegraphics[width=1\columnwidth]{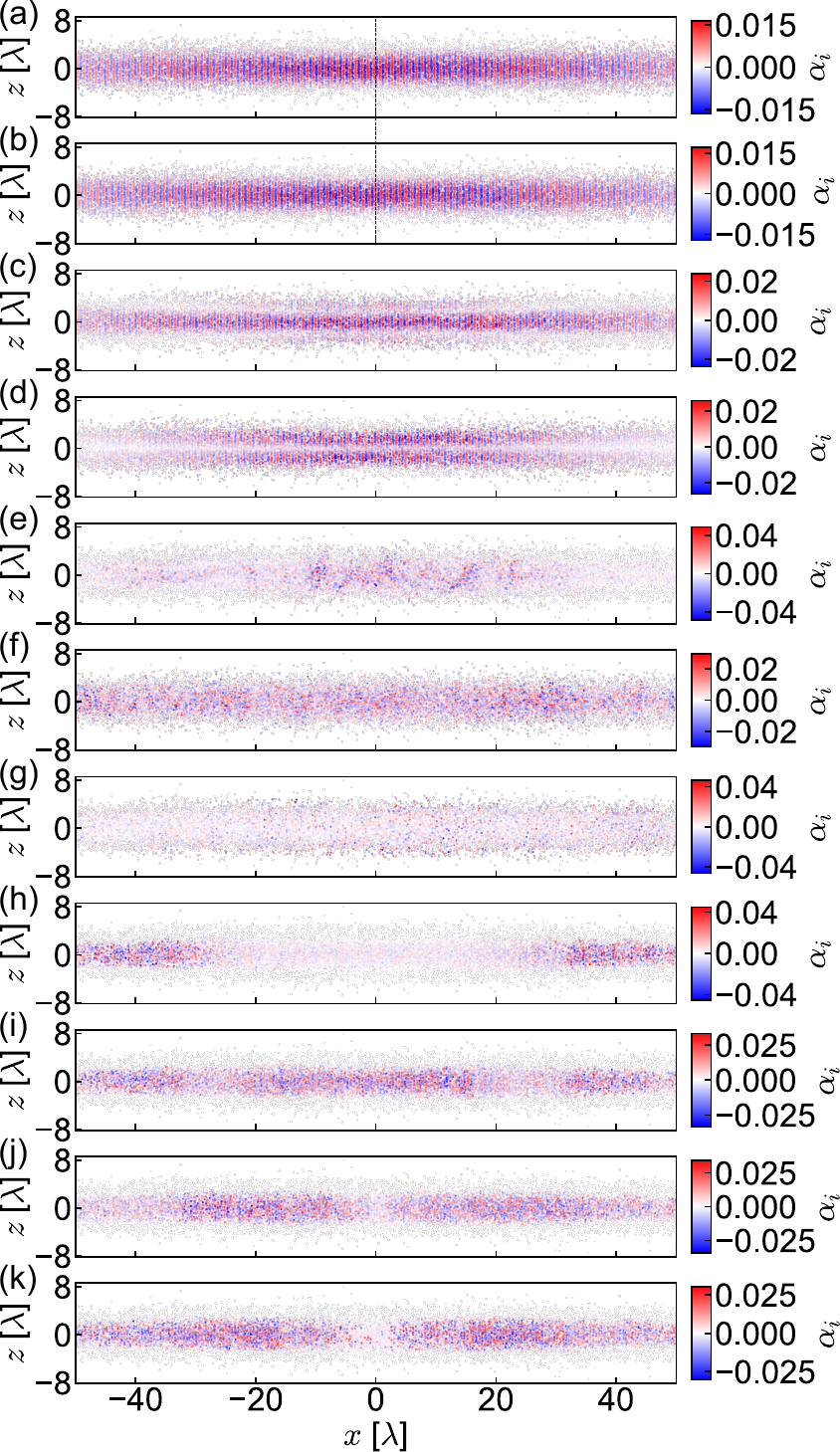}
    \caption{Numerical computation of the eigenvectors of a cloud of $N=30000$ atoms,     for our smallest cloud ($\sigma_x= \SI{33}{\micro m}$, $\sigma_r=\SI{1.4}{\micro m}$). Spatial distributions of $\alpha_{il}$ at positions $\mathbf{r}_i$ for different atomic modes (a) $l=1$, (b) $l=2$, (c) $l=3$, (d) $l=6$, (e) $l=101$, (f) $l=1001$, (g) $l= 15001$, (h) $l=N$, (i) $l=N-2$, (j) $l=N-8$ and (k) $l=N-13$. 
    The atom number used in the simulation ($N=30000$) is  larger than in the experiment: this  does not change the collective mode and spatial dependence but leads to a better fine-grained visualisation. }
    \label{fig:SM-alphas}
\end{figure}

Using the framework presented in the main text, we compute the eigenvectors $S^-_l=\sum_{i}\alpha_{il}\sigma_i^-$ for different modes. Figure~\ref{fig:SM-alphas} shows the spatial distribution of the $\alpha_{il}$ for these modes, and Fig.~\ref{fig:SM-patterns} represent their corresponding radiation patterns $I_l(\mathbf k)$. The polarization of each pattern is given by that of a single 2-level atom with $\mathbf{E}_l(\hat{\mathbf{k}}) \propto \left((\mathbf{\hat{k}} \times \mathbf{p}) \times \mathbf{\hat{k}}  \right) \sum_{i}\alpha_{il}e^{i\mathbf{k}\cdot\mathbf{r}_i}$, with $\mathbf{p}$ the atomic dipole \cite{N-atom-paper}. 
The superradiant atomic modes [Fig.~\ref{fig:SM-alphas}(a-d)] do show a spatial structure and their  radiation patterns [Fig.~\ref{fig:SM-patterns}(a-d)] display a high directionality. However, the emission patterns of the superradiant modes $l\geqslant3$ lie \emph{outside} of the emission of the most superradiant ones ($l=1,2$). 
On the contrary, the subradiant modes [Fig. \ref{fig:SM-alphas}(h-k)] do not feature directional emission pattern as shown in Fig.~\ref{fig:SM-patterns}(e-h). 
We also present the distributions of $\alpha_{il}$ for various modes from the bulk of the eigenvalue distribution in Fig.~\ref{fig:SM-alphas}(e-g): they do not exhibit specific spatial structure. 

\begin{figure}
    \centering
    \includegraphics[width=1\columnwidth]{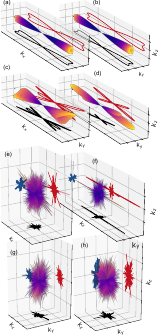}
    \caption{Emission pattern $I_l(\theta,\phi)$  corresponding to the atomic modes $S_l^-$ of Fig.~\ref{fig:SM-alphas} for (a) $l=1$, (b) $l=2$, (c) $l=3$, (d) $l=6$, (e) $l=N$, (f) $l=N-2$, (g) $l=N-8$ and (h) $l=N-13$.}
    \label{fig:SM-patterns}
\end{figure}

\begin{figure}
    \centering
    \includegraphics[width=1\columnwidth]{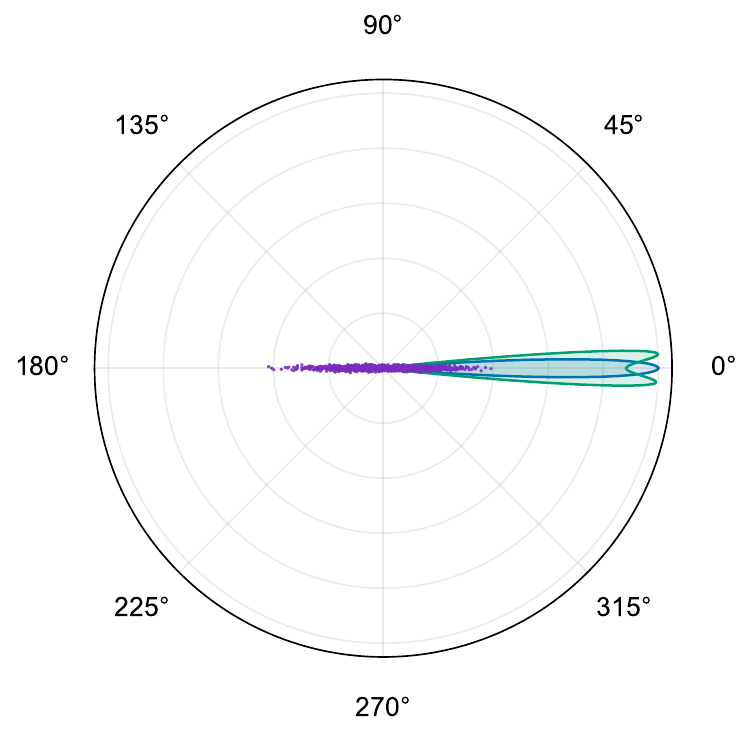}
    \caption{Emission pattern of two single excitation states in the cloud of the form of a spinwave: $\ket\psi=\prod_{\otimes n}(\ket{g_n}+\epsilon e^{ik_\parallel x_n}\ket{e_n})$, where $k_\parallel=k_0$, where $k_\parallel=k_0$ (blue) $k_\parallel=0.997 k_0$ (green). The cloud is our smallest cloud ($\sigma_x= \SI{33}{\micro m}$, $\sigma_r=\SI{1.4}{\micro m}$). }
    \label{fig:pencil_smallk0}
\end{figure}

As mentioned in the text, the cloud exhibits two (nearly-)degenerate most superradiant modes $l=1,2$, with spatial distributions well approximated by $\alpha_{i,l=1}\sim\cos(k_\parallel x_i)$ and $\alpha_{i,l=2}\sim\sin(k_\parallel x_i)$ with $k_\parallel<k_0$. 
Their emission pattern is symmetric along the cloud axis, with the two fields dephased by $\pi/2$. Defining new atomic modes $S_1^-\pm iS_2^-$ leads to a collective, now directional, emission along $\pm x$. 

We finally provide a qualitative explanation of the shape of the spatial pattern of the most superradiant atomic mode.
The total emission power of atomic mode $l$ reads $P_l=\int d\Omega_{\mathbf k}I_l(\mathbf k)$ with $I_l(\mathbf{k})=I_0(\mathbf k)\left|\sum_i \alpha_{il} e^{-i\mathbf{k}\cdot\mathbf{r_i}}\right|^2$.
The spatial pattern of the most superradiant atomic mode results from maximizing the emission power in $4 \pi\,\SI{}{sr}$, i.\,e. maximizing the solid angle in which constructive interferences of the fields radiated by each atom occur. 

To gain an intuition about the shape of the emission pattern we consider the following variational ansatz. We assume the most superradiant mode nearly consists of a spin wave. By symmetry its wavevector is along the cloud axis, $\mathbf{k}_W=k_\parallel \mathbf e_x$ but with $k_\parallel/k_0=\cos \alpha$ a variational parameter. Since we know that the emission occurs in a small angle, we ignore the directional dependence due to the dipole emission pattern. 
We represent in figure \ref{fig:pencil_smallk0} the emission pattern for spinwaves with two different $k_\parallel/k_0$. 
The power emitted by such a spinwave is $P_A\propto \int_{4\pi}d\Omega_\mathbf k\,S(\mathbf k-\mathbf k_W)$, where the structure factor of our Gaussian cloud is $S(\mathbf q)=e^{-\sigma_r^2q_\perp^2-\sigma_x^2q_\parallel^2}$. 
It then reads $P_A\propto 2\pi\int_0^\pi d\theta \sin \theta e^{-k_0^2\sigma_r^2\sin^2\theta}e^{-k_0^2\sigma_x^2(\cos^2\theta-\cos^2\alpha)}$. 
The maximization of $P_A$ then results from the competition of the two first terms: 
The first one reflects the maximization of the solid angle of emission, it favours $\alpha\to\pi/2$. The second one, favouring $\alpha\to0$ expresses the fact that constructive interference cannot take place for too high emission solid angles. 
The shape of the mode thus results from this competition between the finite cloud size and phase matching. 
Numerically maximizing $P_A$, we find a reasonable agreement between this ansatz and the result of the diagonalization of $\Gamma_{nm}$. Our clouds are in the intermediate regime $\sigma_r\gtrsim\lambda_0$: 
In this case superradiance occurs in a cone characterized by $0<k_\parallel<k_0$, as illustrated in Fig.~\ref{fig:SM-patterns}. 
We numerically find that the precise value of the peak emission angle varies with the radial and axial sizes of the cloud.



\begin{figure}
    \centering
    \includegraphics[width=1\columnwidth]{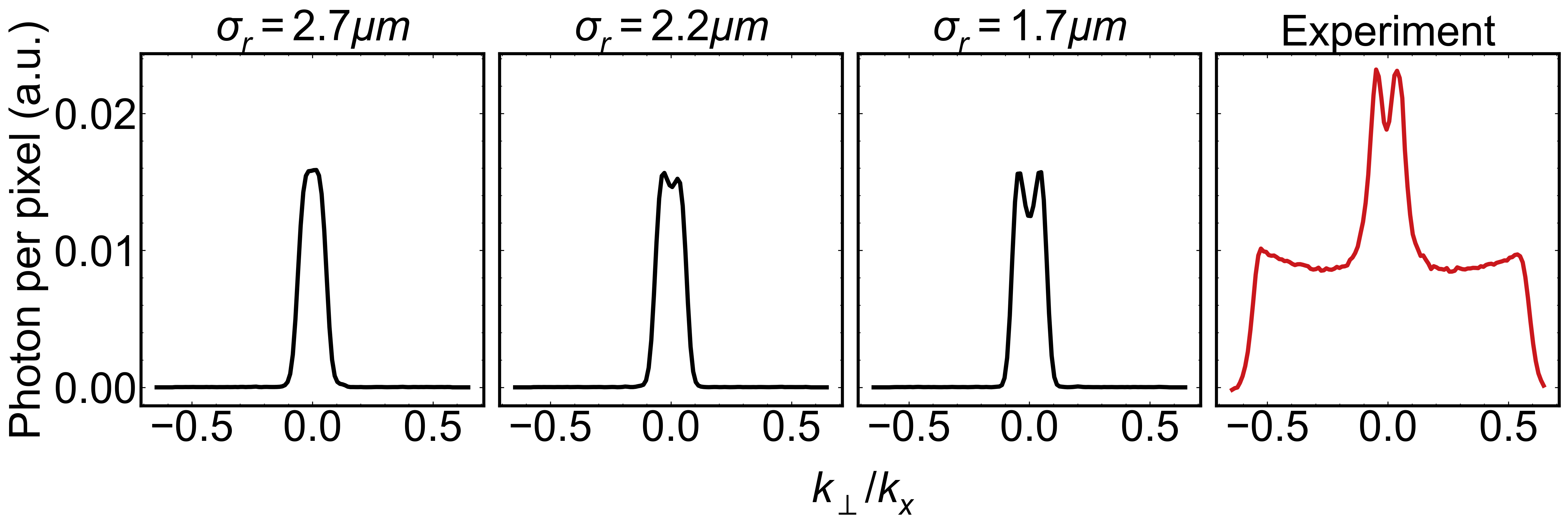}
    \caption{Azimuthal average of the emission pattern of the most superradiant mode, computed for different cloud rms width $\sigma_{r}$ and fixed cloud length $\sigma_{x} = 78\, \mu{\rm m}$. The Simulations amplitudes have been rescaled for comparison, and are thus arbitrary.
    Experimental data  (right) are shown for comparison.}
    \label{fig:SM-patterns-various}
\end{figure}

\section{Condition for superradiance in EM mode $l$}

Following \cite{Robicheaux2021}, we derive the condition for superradiance in EM mode $l$ by calculating the initial slope of the intensity in this mode, starting from a fully inverted system in state $|ee...e\rangle$. 
We use the master equation expressed with the collective atomic modes obtained by diagonalizing the decay matrix $\Gamma_{ij}$, such that 
$\Gamma_{ij} = \sum_l \alpha_{il}\, \Gamma_l\,  \alpha_{jl}$ with $\sum_i \alpha_{il} \alpha_{il'} = \delta_{ll'}$,
and the associated collective jump operators 
$S_l^- = \sum_i \alpha_{il}\, \sigma_i^-$, the $\alpha_{il}$ being real numbers.
We then calculate the mode population $\Pi_l(t)= \langle S_l^+ S_l^-\rangle(t)$ and the corresponding mode intensity $I_l(t) = \Gamma_l\, \Pi_{l}(t)$. 
From the master equation $\dot \rho= \frac{1}{i\hbar}[H_{\rm dd},\rho]+{\cal L}[\rho]$, we get: 
\begin{eqnarray}
    &\dot{\Pi}_{l}(0)& = {1\over i\hbar}\,\langle [H_\text{dd},\, S_l^+ S_l^-]\rangle_0
   \\ &+& \sum_{l'} \Gamma_{l'} \left[
    \langle S_{l'}^+ S_l^+ S_l^- S_{l'}^-\rangle_0
    - \tfrac{1}{2}\langle \{S_{l'}^+ S_{l'}^-,\, S_l^+ S_l^-\}\rangle_0
  \right] \nonumber.
  \label{eq:Cll_dot}
\end{eqnarray}
where $\langle O \rangle_0 = \text{Tr}\left(O\rho(t=0)\right)$.

Here, we have included in the master equation the Hamiltonian term $H_\text{dd}=\sum_{i\neq j} J_{ij}(\sigma_i^+\sigma_j^-+\sigma_i^-\sigma_j^+)$ describing the resonant dipole interactions, with $J_{ij}\propto \text{Re}[G\left(\mathbf{r}_i-\mathbf{r}_j\right)]$ \cite{Lehmberg1970}. This term was ignored in the main text.  
However, here it does not need to be ignored as the trace of the $J_{ij}$ matrix is 0 since an atom does not interact with itself. 
The evaluation of the anticommutator gives $\tfrac{1}{2} \langle \{S_{l'}^+ S_{l'}^-,\, S_l^+ S_l^-\}\rangle_0 = 1$, 
using $\sum_i \alpha_{il}^2=1$.
For the first term, we write $\langle S_{l'}^+ S_l^+ S_l^- S_{l'}^-\rangle_0  = \| S_l^- S_{l'}^-\, |e\cdots e\rangle \|^2$ and calculate 
\begin{equation}
  S_l^- S_{l'}^-\, |e\cdots e\rangle
  = \sum_{a\neq b} \alpha_{al'} \alpha_{bl}\, |g_a, g_b\rangle,
\end{equation}
where $|g_a,g_b\rangle$ denotes the state with atoms $a$ and $b$ in their ground state. 
Using the symmetry under $a\leftrightarrow b$, we get
\begin{align}
  \| S_l^- S_{l'}^-\, |e\cdots e\rangle\|^2
  &= \sum_{a<b}(\alpha_{al'} \alpha_{bl} + \alpha_{bl'} \alpha_{al})^2 \nonumber\\
  &= \sum_{a\neq b}\alpha_{al'}^2\alpha_{bl}^2
   + \sum_{a\neq b} \alpha_{al'} \alpha_{al}\, \alpha_{bl} \alpha_{bl'}.
\end{align}
Introducing the mode-overlap:
$  \mathcal{O}_{ll'} = \sum_a \alpha_{al}^2\alpha_{al'}^2$,
and using the eigenvector orthonormality, the two sums yield
\begin{align}
  \sum_{a\neq b}|\alpha_{al'}|^2|\alpha_{bl}|^2 &= 1 - \mathcal{O}_{ll'}, \\
  \sum_{a\neq b} \alpha_{al'} \alpha_{al}\, \alpha_{bl} \alpha_{bl'}
   &= \delta_{ll'} - \mathcal{O}_{ll'}.
\end{align}
so that
\begin{equation}
  \| S_l^- S_{l'}^-\,|e\cdots e\rangle \|^2 = 1 + \delta_{ll'} - 2\mathcal{O}_{ll'}.
  \label{eq:recycling}
\end{equation}
Substituting equation \eqref{eq:recycling} and the anticommutator value
into equation \eqref{eq:Cll_dot} leads to
\begin{equation}
  \dot\Pi_{l}(0) = \sum_{l'} \Gamma_{l'}
  \left[ \delta_{ll'} - 2\mathcal{O}_{ll'} \right].
\end{equation}
As $\sum_{l'} \Gamma_{l'}|\alpha_{al'}|^2 = \Gamma$, and $\sum_a |\alpha_{al}|^2 = 1$, we obtain $\sum_{l'}\Gamma_{l'}\mathcal{O}_{ll'} = \Gamma$, yielding the expression
\begin{equation}
    \dot{I}_l(0) = \Gamma_l\,(\Gamma_l - 2\Gamma_0).
  \label{eq:Il_dot_result}
\end{equation}
This gives the condition for superradiance in EM mode $l$:
\begin{eqnarray}
    \Gamma_l>2\Gamma_0 
\end{eqnarray}
This condition should be contrasted to the ones obtained in Refs \cite{Masson2022,Robicheaux2021} where the criterion is derived in $4\pi$ or \emph{directionally}. 
Since the collective modes have a decay $\Gamma_l$ proportional to $N$, this criterion defines a critical atom number \emph{per mode} $N_{{\rm c},l}$ defined by $\Gamma_l(N_{{\rm c},l})=2\Gamma_0$. 

\begin{figure}
    \centering
    \includegraphics[width=1\columnwidth]{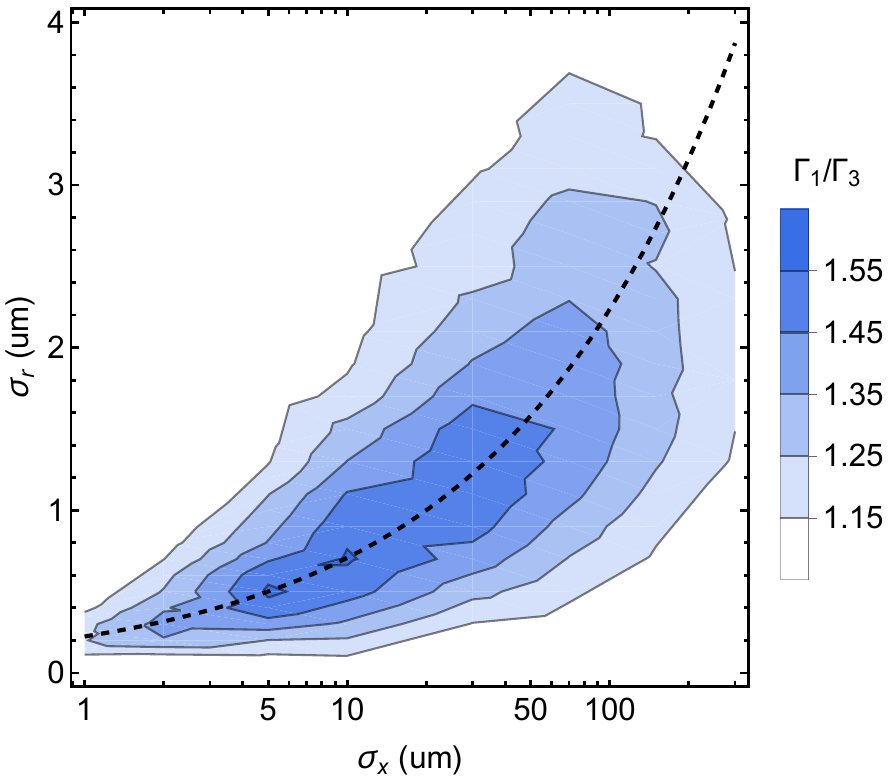}
    \caption{Amplitude of the spectral gap $\Gamma_1/\Gamma_3$ as a function of the cloud length $\sigma_x$ and radius $\sigma_r$, for $N=6000$. We find that this ratio is independent of $N$ for large $N$. Dashed line: guide to the eye corresponding to constant Fresnel number ${\cal F}=\frac{\pi\sigma_r^2}{\lambda_0\sigma_x}=0.2$.}
    \label{fig:gap_figure}
\end{figure}

\subsection{Spectral gap in the eigenvalue distribution}

The distribution of eigenvalues $\Gamma_l$ of the $\Gamma_{ij}$ matrix has been discussed in the context of random matrix theory \cite{Skipetrov2011,Viggiano2023}. 
Most studies focus on the statistical properties of the distribution of eigenvalues and their spacings. 
Here, we are interested in the existence of a gap between the most superradiant mode $l=1,2$ and the next one $l=3$. 

For our cloud sizes, we do find a gap, as is visible in Figure \ref{fig:modes}. 
We plot in Fig.~\ref{fig:gap_figure} the magnitude of this gap, defined by the ratio between the two first eigenvalues pairs $\Gamma_1/\Gamma_3$, as a function of cloud length $\sigma_x$ and cloud radius $\sigma_r$, with $N=6000$ atoms. We numerically observe a significant gap, $\Gamma_1/\Gamma_3>1$, for a large range of cloud sizes $(\sigma_x,\sigma_r)$.
We further find that the largest gap is reached along a curve (dashed line) defined by a constant Fresnel number ${\cal F}=\pi\sigma_r^2/(\lambda_0\sigma_x)\approx0.2$ \cite{Gross1982}. 
Our experiments lie in the range $\sigma_{x} = 20-80\,\mu{\rm m}$, $\sigma_r = 1-2.5\,\mu{\rm m}$ where a gap does exist. 
Since the existence of this gap is necessary for one single mode to dominate the superradiant decay, as explained in the main text, it would be interesting to investigate further its condition of existence in relation to the properties of the $\Gamma_{ij}$ matrix.

\section{Limitations of the bosonic mode approximation}

The bosonic approximation described in the main text is only exact in the Dicke limit where one collective atomic mode, and hence one spontaneous decay channel, plays a role \cite{Clemens2003}. 
In our extended sample we thus do not expect it to reproduce fully our observations. 
Several features of superradiance are however qualitatively predicted by the model: the burst in intensity, shown in Fig.~\ref{fig:bosonlimit}, and the amplification of the most superradiant atomic mode. 
Nonetheless, the peak intensity 
grows faster with atom number ($>N^3$) than observed experimentally, and the duration of the burst is longer than the experimental one by a factor $\sim 2$, as shown in Fig.~\ref{fig:bosonlimit}. 
The discrepancy grows with atom number, and superradiance is thus overestimated by the model.
It also does not predict a threshold in atom number for the superradiance to occur. 
Furthermore, we observe numerically that the model never predicts an exponential decay of the intensity, even in the case of independent atoms.  
This can be understood by the fact that even for independent emitters ($\Gamma_l = \Gamma_0$ for all $l$), the bosonic amplification remains and allows for multiple jumps by the same atom after a first photon emission. 

\begin{figure}
    \centering
    \includegraphics[width=1\columnwidth]{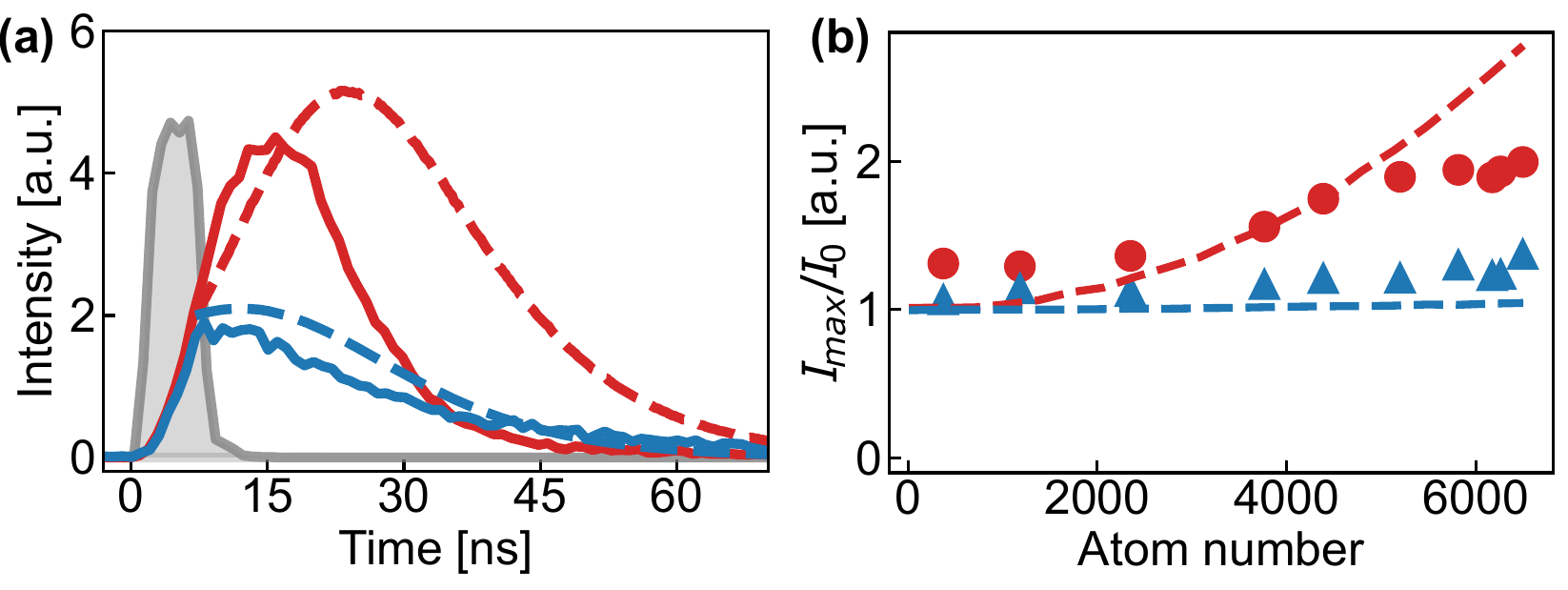}
    \caption{(a) Superradiant burst measured along (red) and orthogonal (blue) to the cloud's axis. 
    (b) Scaling of the maximum intensity relative to the initial intensity for various number of atoms. 
    (a) and (b) The dashed lines represent the prediction of the boson approximation model. The solid lines, circle and triangles are the experimental data}
    \label{fig:bosonlimit}
\end{figure}

\end{document}